\documentclass[aps,pre,10pt,showpacs,floatfix,onecolumn,nofootinbib,a4paper]{revtex4-2}

\usepackage{amssymb,amsmath,mathtools}
\usepackage{bm}
\usepackage{mathrsfs}
\usepackage{hyperref,bookmark}
\usepackage{graphicx}
\usepackage{epsfig,color}
\usepackage{float}
\usepackage{subcaption}
\usepackage{verbatim}
\usepackage{dcolumn}
\usepackage{tikz}
\usepackage{url}
\usepackage{stmaryrd}

\usepackage[utf8x]{inputenc}
\usepackage[english]{babel}
\usepackage{mathrsfs}
\usepackage{epsfig}
\usepackage{listings}
\usepackage{paralist}

\usepackage{mdwlist}
\usepackage{dsfont}
\usepackage{oldgerm}
\usepackage{geometry}
\usepackage{relsize}
\usepackage{afterpage}
\usepackage{lmodern}
\usepackage{tabularx}
\usepackage{cancel}

\newcommand{\dd}{\mathrm{d}}

\newcommand{\vt}{\theta}
\newcommand{\vp}{\phi}
\newcommand{\trans}{^\mathsf{T}}

\newcommand{\n}{\bm{n}}
\newcommand{\m}{\bm{m}}
\newcommand{\nper}{\bm{n}_\perp}
\newcommand{\mper}{\bm{m}_\perp}
\newcommand{\p}{\bm{p}}
\newcommand{\cv}{\bm{c}}

\newcommand{\e}{\bm{e}}

\newcommand{\zero}{\bm{0}}
\newcommand{\y}{\bm{y}}

\newcommand{\normal}{\bm{\nu}}
\newcommand{\surface}{\mathscr{S}}

\newcommand{\free}{\mathscr{F}}

\newcommand{\nablas}{\nabla\!_\mathrm{s}}

\newcommand{\W}{\mathbf{W}}
\newcommand{\F}{\mathbf{F}}

\newcommand{\I}{\mathbf{I}}

\newcommand{\R}{\mathbf{R}}
\newcommand{\C}{\mathbf{C}}

\newcommand{\U}{\mathbf{U}}

\newcommand{\tr}{\operatorname{tr}}

\newcommand{\curvature}{(\nablas\normal)}

\newcommand{\euclid}{\mathscr{E}}
\newcommand{\curve}{\mathscr{C}}

\renewcommand{\frame}{(\e_1,\e_2,\e_3)}
\newcommand{\framem}{(\m,\mper,\e_3)}

\newcommand{\framen}{(\n,\nper,\normal)}

\newcommand{\tangent}{\bm{t}}
\newcommand{\tangentper}{\bm{t}_\perp}

\newcommand{\ridge}{\bm{r}}

\renewcommand{\leq}{\leqslant}
\renewcommand{\geq}{\geqslant}

\begin{document}
	\title{Ridge approximation for thin  nematic polymer networks}
	\author{Andrea Pedrini}
	\email{andrea.pedrini@unipv.it}
	\author{Epifanio G. Virga}
	\email{eg.virga@unipv.it}
	\affiliation{Dipartimento di Matematica, Universit\`a di Pavia, Via Ferrata 5, 27100 Pavia, Italy }

	\date{\today}

\begin{abstract}
Nematic polymer networks (NPNs) are nematic elastomers within which the nematic director is enslaved to the elastic deformation. The elastic free energy of a NPN sheet of thickness $h$ has both stretching and bending components (the former scaling like $h$, the latter scaling like $h^3$). NPN sheets bear a director field $\m$ imprinted in them (usually, uniformly throughout their thickness); they can be activated by changing the nematic order (e.g. by illumination or heating). This paper illustrates an attempt to compute the bending energy of a NPN sheet and to show which role it can play in determining the activated shape. Our approach is approximate: the activated surface  consists of  flat sectors connected by \emph{ridges}, where the unit normal jumps and  the bending energy is concentrated. By increasing the number of ridges, we should get closer to the real situation, where the activated surface is smooth and the bending energy is distributed on it.
%(at this stage, however, this is more an expectation than an established fact).
The method is applied to a disk with imprinted a spiraling hedgehog. It is shown that  upon activation the disk, like a tiny  hand, is able to grab a rigid lamina.  
\end{abstract}
	
\maketitle

\section{Introduction}\label{sec:intro}
Nematic polymer networks (NPNs) are elastomeric networks whose polymeric chains have a nematogenic component that is so tightly connected with the polymeric network as to enslave the nematic director to the elastic deformation. Thus, when the scalar nematic order is altered by an external stimulus (such as heat of light), a thin NPN sheet is driven out of equilibrium and its shape changes. The deformed equilibrium configuration is determined by minimizing the elastic free energy of the system \cite{nguyen:theory}.
As customary in plate theory, this latter has a stretching component, which scales like the thickness $h$ of the sheet, and a bending component, which scales like $h^3$ \cite{ozenda:blend}. 

Minimizing the stretching component amounts to prescribe a metric, which is determined by the nematic director field $\m$ imprinted in the flat, undeformed sheet (generally, uniformly across its thickness) \cite{modes:disclination,modes:gaussian}. Often, with a slight abuse of language, the deformation realizing the prescribed metric is called an \emph{isometric immersion}.\footnote{The abuse is pardoned by thinking of the flat, undeformed configuration as being endowed with the prescribed metric, which is kept by the immersion in three dimensional space.} By Gauss' \emph{theorema egregium}, prescribing a metric also means prescribing the Gaussian curvature of the immersed surface. Due to this constraint, the bending energy is, as it were, less effective on isometric immersions than it would be in general. 

The problem of determining the isometric immersions dictated by a given imprinted field $\m$ is simple to state, fascinating, but hard \cite{aharoni:geometry,mostajeran:curvature,mostajeran:encoding}, despite its neglect of the bending energy. Clearly, incorporating the latter would make the problem even harder.

Here we wish to move a step toward solving this harder problem, where the activated shape of a NPN thin sheet is determined by both the stretching and bending components of the elastic free energy. We adopt a two-step strategy: minimizing the dominant stretching energy first and then let the bending energy act as a selection criterion on the minimizers of the stretching energy. 

As explained in Sec.~\ref{sec:energies}, our strategy relies on an approximation: we use piece-wise isometric immersions with flat sectors connected along \emph{ridges}, lines of discontinuity for the immersed surface. The bending energy is concentrated along ridges; increasing their number should provide a better approximation of the real sheet with its surface bending energy distribution.\footnote{Unfortunately, we do not yet possess a formal convergence result; we shall be contented with considering a sufficiently large number of ridges.} 
In Sec.~\ref{sec:strategy}, we illustrate our general strategy and give conditions under which it can be applied. An application is given in Sec.~\ref{sec:hedgehogs}, where we determine numerically the activated shape of a disk with imprinted a spiraling hedgehog when it grabs a rigid lamina surmounting it. Finally, in Sec.~\ref{sec:conclusions}, we recap our conclusions and comment on other possible applications of our approximate method. The paper is closed by two appendices: in one, we give the mathematical details needed to follow our development in Sec.~\ref{sec:strategy}; in the other, we illustrate the algorithm used to minimize the total ridge energy for the problem studied in Sec.~\ref{sec:hedgehogs}. 

\section{Stretching, bending, and ridge energies}\label{sec:energies}
A thin NPN sheet will be treated here as a flexible, inextensible material surface, which in the absence of external stimuli takes on a flat reference shape $S$ in the $(x_1,x_2)$ plane of a Cartesian frame $\frame$. Upon activation, it generally becomes a curved surface $\surface$ in space, which is the image of $S$ under the deformation $\y$, $\surface=\y(S)$. A director field $\m$ is imprinted on $S$, it represents the average orientation of the nematic molecules entrapped in the polymer matrix at the time of crosslinking \cite{nguyen:theory}.

In NPNs, unlike general nematic elastomers, the link between nematic molecules and polymer matrix is so strong that the latter entrains the former in its deformation \cite{white:programmable}. Only the scalar order parameter can be affected by external stimuli (such as heat and light), it plays the role of an activation parameter in our theory and will be denoted by $s_0$ in the reference configuration and by $s$ in the present configuration. It is precisely the deviation of $s$ from $s_0$ that drives a thin NPN sheet  out of equilibrium, causing it to deform.

\begin{figure}[h]
	\begin{center}
		\includegraphics[width=.4\linewidth]{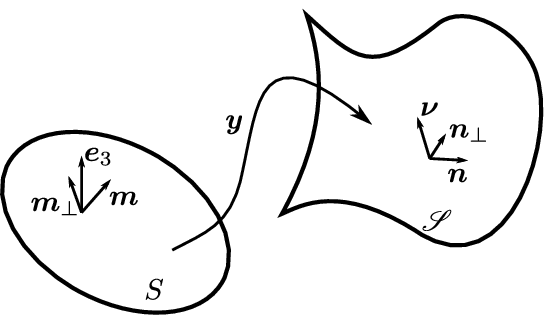}
	\end{center}
	\caption{\label{fig:sketch} A flat surface $S$ in the $(x_1,x_2)$ plane of a fixed Cartesian frame $\frame$ is deformed by the mapping $\y$ into a smooth surface $\surface$ embedded in three-dimensional Euclidean space $\euclid$. The blueprinted orientation is denoted by $\m$ in the reference configuration and by $\n$ in the current one; $\e_3$ is the outer unit normal to $S$, while $\normal$ is the outer unit normal to $\surface$;  correspondingly, $\mper:=\e_3\times\m$ and $\nper:=\normal\times\n$.}
\end{figure}

Figure~\ref{fig:sketch} depicts the deformation of $S$.
The frame $\framem$ is oriented like the frame $\frame$, with $\e_3$ normal to $S$. The deformation $\y$ takes $\framem$ into the frame $\framen$, where $(\n,\nper)$ lies on the local tangent plane of $\surface$ and $\normal$ is the unit normal to $\surface$, directed so as to make the orientation of $\framen$ agree with that of $\framem$. The strong coupling between nematic molecules and polymer network makes $\n$ the nematic director in the current configuration $\surface$; $\n$ is formally delivered by
\begin{equation}
	\label{eq:n_y_coupling}
	\n=\frac{(\nabla\y)\m}{|(\nabla\y)\m|}.
\end{equation}

The elastic free energy stored in $\surface$ as the result of a deformation $\y$ was derived in \cite{ozenda:blend} via a dimensional reduction based on a revised Kirchhoff-Love hypothesis \cite{ozenda:kirchhoff} from the acclaimed \emph{trace formula} of Terentjev and Warner for the elastic free energy in three space dimensions (see Chap.~6 of \cite{warner:liquid}). As customary in plate theory, the elastic free energy $f_e$ stored per unit area of a thin sheet can be expanded in odd powers of (half) its thickness $h$,
\begin{equation}\label{eq:free_energy_expansion}
	f_e=hf_1+h^3f_3+O(h^5),
\end{equation}
where $f_s:=hf_1$ is the \emph{stretching} energy and $f_b:=h^3f_3$ is the \emph{bending} energy.\footnote{Here $f_e$ is scaled to an elastic modulus having physical dimensions of an energy per unit volume, see \cite{pedrini:ridge}.} These were found to be 
\begin{subequations}
	\begin{eqnarray}
		f_s&=&\frac{2h}{s+1}\left(\tr\C+s_0\m\cdot\C\m+\frac{s}{\m\cdot\C\m}\right),\label{eq:f_s}\\
		f_b&=&\frac{2h^3}{3}\left\{2(8H^2-K)+\frac{1}{s+1}\left[\left(\frac{3s}{a^2}-a^2s_0-\tr\C\right)K-\frac{4s}{a^2}(2H-\kappa_n)\kappa_n\right] \right\},\label{eq:f_b}
	\end{eqnarray}
\end{subequations}
respectively, where $\C:=(\nabla\y)\trans(\nabla\y)$ is the \emph{right} Cauchy-Green tensor, which measures how lengths and angles on $S$ changes upon deformation; $\C$ is the \emph{metric} tensor associated with the deformation $\y$. The eigenvalues of $\C$ are the squared principal stretches, $\lambda_1^2$ and $\lambda_2^2$; their product is constrained by the requirement
\begin{equation}
	\label{eq:det_C_constrained}
	\lambda_1^2\lambda_2^2=\det\C=1,
\end{equation}
which expresses the inextensibility of $\surface$. $H$ and $K$ are the mean and Gaussian curvatures of $\surface$, defined as
\begin{equation}
	\label{eq:H_K_definitions}
	H:=\frac12\tr\curvature\quad\text{and}\quad K:=\det\curvature
\end{equation}
in terms  of the curvature tensor $\nablas\normal$. Finally,
\begin{equation}
	\label{eq:a^2_kappa_nn}
	a^2:=\m\cdot\C\m\quad\text{and}\quad\kappa_n:=\n\cdot\curvature\n.
\end{equation}

Within the plate approximation, the total elastic free energy is then the functional defined by
\begin{equation}
	\label{eq:free_energy_plate}
	\free[\y]:=\int_S(f_s+f_b)\dd A,
\end{equation}
where $\dd A$ is the area element. For prescribed $s\neq s_0$, the minimizers of $\free$ subject to \eqref{eq:det_C_constrained} describe the activated shapes of the sheet. As made apparent by \eqref{eq:f_b}, both stretching and bending measures are mixed in $f_b$, whereas $f_s$ depends only on the principal stretches. It is precisely the blend of stretching and bending in \eqref{eq:free_energy_plate} that makes it hard to minimize $\free$, even under the simplest boundary conditions \cite{pedrini:ridge}.

The most common approach to this optimization problem has been to neglect $f_b$ in $\free$ and minimize $f_s$ (see, for example, \cite{modes:blueprinting,modes:gaussian,modes:negative,mostajeran:curvature,mostajeran:encoding,mostajeran:frame}). A simple computation \cite{ozenda:blend} shows that $f_s$ is minimized for
\begin{equation}\label{eq:C_stationary}
	\C=\C_0:=\lambda_1^2\m\otimes\m+\lambda_2^2\mper\otimes\mper,
\end{equation}
where, in agreement with \eqref{eq:det_C_constrained}, 
\begin{equation}\label{eq:lambda_1_2}
	\lambda_1:=\sqrt[4]{\frac{s+1}{s_0+1}}\quad\text{and}\quad\lambda_2=\frac1\lambda_1.
\end{equation}
The principal stretches delivered by \eqref{eq:lambda_1_2} carry a transparent physical meaning. When $s>s_0$, which is the case upon cooling the material so that the degree of molecular order is increased, $\lambda_1>1$ and $\lambda_2<1$, meaning that the material fibers are extended along $\m$ and shortened in the orthogonal direction $\mper$. Clearly, the converse behavior is anticipated for $s<s_0$. A deformation $\y$ for which \eqref{eq:C_stationary} holds is called an \emph{isometric immersion}.

As shown in \cite{mostajeran:curvature} (see also equation (43) of \cite{ozenda:blend} for an equivalent formulation), the deformed surface $\surface$ produced by an isometric immersion has Gaussian curvature determined by the imprinted director field $\m$,
\begin{equation}\label{eq:theorema_egregium}
	K=\left(\lambda_1^2-\lambda_2^2\right)(c_2^2-c_1^2+c_{12}),
\end{equation}
where
\begin{equation}
	\label{eq:connector_components}
	c_1:=\cv\cdot\m,\quad c_2:=\cv\cdot\mper,\quad c_{12}:=\m\cdot(\nabla\cv)\mper,
\end{equation}
and $\cv$ is the \emph{connector} of $\m$, defined as $\cv:=(\nabla\m)\trans\m$. Equation \eqref{eq:theorema_egregium} is the incarnation in the present context of a  theorem in differential geometry of surfaces, Gauss' \emph{theorema egregium} \cite[p.\,144]{stoker:differential}. Thus, for an isometric immersion only $H$ remains as an independent measure of bending.

Despite the large body of elegant literature that has been produced by approximating $f_e$ with $f_s$  (the list \cite{modes:gaussian,aharoni:geometry,modes:negative,plucinsky:programming,mostajeran:curvature,mostajeran:encoding,mostajeran:frame,kowalski:curvature,warner:nematic,aharoni:universal} is just a representative sample), a number of issues have arisen. First, there is no guarantee that an isometric immersion exists in the large for arbitrary choices of the imprinted field $\m$ (whereas, of course, it always exists locally). If no global isometric immersion exists, this approximation method clearly fails. But it also fails if there are many, as they all have the same stretching energy. This is the case, for example, for the axisymmetric surfaces produced by lifting a field $\m$ imprinted on a flat disk \cite{modes:gaussian,mostajeran:encoding}. The natural selection criterion in these cases was provided by minimizing an \emph{ad hoc} additional bending energy, taken proportional to $H^2$, which is but a special case of $f_b$ in \eqref{eq:f_b}.

In \cite{pedrini:ridge}, we took a different approach, on which here we wish to build an alternative approximation. We considered piece-wise smooth immersions that map $S$ into a \emph{ridged} surface $\surface$, that is, a surface possibly traversed by lines along which the outer normal $\normal$ suffers a jump, while the nematic director $\n$ stays continuous.

\begin{figure}[h]
	\centering
	\includegraphics[width=.4\textwidth]{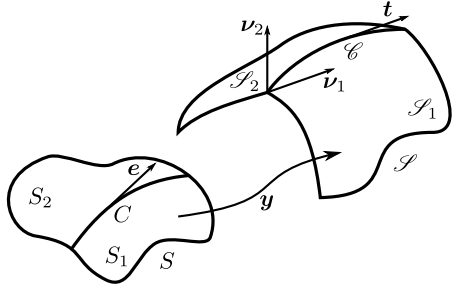}
	\caption{The reference surface $S$ is split by a smooth curve in two sides, $S_1$ and $S_2$, which a deformation $\y$, continuous through $C$, but with discontinuous gradient, maps into the sides $\surface_1$ and $\surface_2$ of the ridge $\curve$.}
	\label{fig:ridge}
\end{figure}

The generic situation we envision is sketched in Fig.~\ref{fig:ridge}, where $\curve$, which is the image under $\y$ of a line $C$ on $S$, is a \emph{ridge} of $\surface$ bordering on $\surface_1$ and $\surface_2$, images of $S_1$ and $S_2$, respectively, the components that split $S$.
We denote by $\tangent$ the unit tangent to $\curve$; it is related to the unit tangent $\e$ to $C$ as $\n$ is related to $\m$ in \eqref{eq:n_y_coupling} (see \cite{pedrini:ridge}, for more details).

In \cite{pedrini:ridge}, we argued that a bending energy is to be associated with $\curve$, extracted, as it were, from a fold connecting $\surface_1$ and $\surface_2$ at a length scale comparable with $h$. By use of \eqref{eq:f_b}, it was found that such a \emph{ridge energy} has a density (per unit length) given by
\begin{equation}
	\label{eq:ridge_energy_density}
	f_r:=\frac83h^2\arccos^2(\normal_1\cdot\normal_2)\left(1-\frac{s}{s+1}\frac{1}{a^2}(\n\cdot\tangent)^2(\n\cdot\tangentper)^2\right).
\end{equation}  
Since in our theory $f_r$ scales like $h^2$, it can be taken as the leading correction to $f_s$ in $f_e$, so that \eqref{eq:free_energy_expansion} is replaced by
\begin{equation}\label{eq:free_energy_approximation}
f_e=hf_1+h^2f_2+O(h^3),	
\end{equation}
where $hf_1=f_s$ and $h^2f_2=f_r$. Thus, strictly speaking, on the smooth components of $\surface$ the bending energy $f_b$ becomes a higher order perturbation, which can be neglected, so that in the exemplary situation depicted in Fig.~\ref{fig:ridge} we could possibly take both $\surface_1$ and $\surface_2$ as flat and $\curve$ as a straight segment. 

In the following section, we shall base on \eqref{eq:free_energy_approximation} our strategy to describe the bending of a thin NPN sheet.

\section{Ridge strategy}\label{sec:strategy}
\begin{figure}[h]
	\centering
	\includegraphics[width=.4\textwidth]{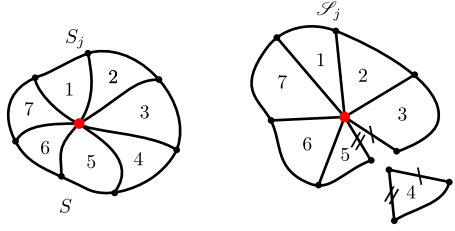}
	\caption{Sketch of a possible decomposition of $S$ in curved triangles $S_j$ that are mapped by isometries satisfying \eqref{eq:C_stationary} into planar triangles $\surface_j$ with two straight sides. The hub is the red point whence all proto-ridges emanates. Adjacent sides have the same length, but all $\surface_j$'s do not fit together. For example, there would not be enough room between $\surface_3$ and $\surface_5$ to accommodate $\surface_4$, although the inner sides of the latter match those equally marked in the former two. Here, all $\surface_j$'s, but $\surface_4$ have been glued together by adjusting the $\y_j$'s with appropriate rigid motions. Only one $\surface_j$ does not fit; in general, there could be more.}
	\label{fig:decomposition}
\end{figure}

The idea at the basis of our strategy is finding a decomposition (with curved triangles) of $S$ that satisfies the following requirements: 
\begin{inparaenum}[(1)]
\item the components $\{S_j\}_{j=1,\dots,N}$ that decompose $S$ are subject to deformations $\y_j$ that change each $S_j$ into a planar domain $\surface_j$ so as to satisfy \eqref{eq:C_stationary} everywhere;\label{item_1} 
\item the sides within $\surface$ of $\surface_j$ (which are the images of the sides within $S$ of $S_j$) are straight segments.\label{item_2} 	
\end{inparaenum}
Figure~\ref{fig:decomposition} shows a possible example of the decomposition we have in mind.
The adjacent sides of two consecutive $\surface_j$'s have the same length because the metric tensors $\C_j=(\nabla\y_j)\trans(\nabla\y_j)$ are one and the same $\C_0$. Of course, since in general $\C_0$ is not the (two-dimensional) identity, putting together in the plane all the $\surface_j$'s by gluing all their adjacent inner sides does \emph{not} deliver a shape $\surface$ that could be regarded as an isometric planar immersion of $S$. Either two sides fall short of one another so that the gap between them cannot be filled, or they come too close to one another so that two consecutive $\surface_j$'s would glide one on top of the other, should they remain on the same plane. Figure~\ref{fig:decomposition} illustrates the latter circumstance, the former can be easily imagined.

Since all $\surface_j$'s cannot be accommodated in the plane, they must be lifted in space. This can be done by viewing the straight sides of the $\surface_j$'s as ridges of the deformed image  of $S$ in space. To make this strategy viable, we must ensure that both requirements can indeed be met. To the accomplishment of these tasks we devote Appendix~\ref{sec:tasks}; here we just collect the answers provided by that analysis. 

We start with the first. By Gauss' \emph{theorema egregium}, since each $\surface_j$ is flat, for a $\y_j$ to exist the imprinted field $\m$ must be such that $K$ in \eqref{eq:theorema_egregium} vanishes identically. It is shown in Sec.~\ref{sec:first_task} that such a necessary condition is indeed also sufficient.

Our second task is finding arcs in $S$ that may constitute the inner boundaries (within $S$) of the $S_j$'s, that is, arcs that  the planar isometries  $\y_j$'s map into straight segments. Letting the unit tangent vector $\e$ to such a generic curve $C$ be represented in the local basis $(\m,\mper)$ as
\begin{equation}\label{eq:e_representation}
\e=\cos\vp\m+\sin\vp\mper
\end{equation}	
and denoting by $\kappa$ its curvature, we prove in Sec.~\ref{sec:second_task} that  $C$ must obey by the equation
\begin{equation}
	\label{eq:protoridge_equation}
	(\lambda_1^2-\lambda_2^2)(c_2\lambda_2^2\sin^3\vp-c_1\lambda_1^2\cos^3\vp)=\lambda_1^2\lambda_2^2\kappa,
\end{equation} 
which is effectively a first order differential equation for $\vp$. It determines a curve $C$, once $\e$ is given at an initial point.\footnote{The reader may observe that for $\lambda_1=\lambda_2$ all solutions to \eqref{eq:protoridge_equation} are straight segments.} The name given to such a curve in \cite{mostajeran:frame} was ``proto-radius'', as it is the pre-image of a radius of $\surface$. We shall prefer calling them \emph{proto-ridges}. Choosing a \emph{hub}  in $S$ (like the red point  depicted in Fig.~\ref{fig:decomposition}) and letting $\e$ take different orientations there, we may obtain a full range of proto-ridges that traverse $S$.  

One may ask under what general circumstances all proto-ridges are straight segments (apart from when $\lambda_1=\lambda_2$). The answer provided by \eqref{eq:protoridge_equation} is: Whenever the curve such that its tangent makes the angle
\begin{equation}
	\label{eq:phi_0}
	\vp=\arctan\left(\sqrt[3]{\frac{\lambda_1^2c_1}{\lambda_2^2c_2}}\right)
\end{equation} 
with $\m$ is a straight line, admittedly a rather peculiar situation.

Our strategy will be finding as many proto-ridges we can handle so as to produce an approximation of $\surface$ by lifting in space its ridges, which cannot be accommodated in a single plane. The advantage is clearly the simplicity of the energy to be minimized. Since all components $S_j$ of $S$ are stretched in the optimal way (i.e., by minimizing $f_s$), we need only minimize the total ridge energy $\free_r$, which is obtained by adding together the energies of all ridges,
\begin{equation}
	\label{eq:total_ridge_energy}
	\free_r:=\sum_{j=1}^N\int_{\curve_j}f_r\dd\ell,
\end{equation}
where $f_r$ is as in \eqref{eq:ridge_energy_density} and $\dd\ell$ is the length element.

We do not have (yet) the luxury of a convergence theorem, which might tell how many ridges should $\surface$ possess for its ridge energy to come close to the parent bending energy we are approximating. We only expect that the approximation would improve upon increasing $N$, but we must rely to some degree on experimenting with $N$.

In the following section, we shall see examples for which the strategy outlined here is indeed successful.

\section{Grabbing a lamina with a hedgehog}\label{sec:hedgehogs}
Here we put our strategy to the test. We first show that both requirements  laid down in the preceding section are satisfied by a family of fields $\m$ imprinted on a disk of radius $R$. This family, which has already been studied in \cite{mostajeran:frame} with different tools (and also explored experimentally in \cite{dehaan:engineering,ware:programmable}), is constituted by planar \emph{spiraling hedgehogs} that interpolate between the pure-splay radial hedgehog and the pure-bend circular field. Once our strategy is proven legitimate, by minimizing numerically the functional $\free_r$ in \eqref{eq:total_ridge_energy}, we shall find the lifted shape of the activated disk able to grab a rigid lamina.

\subsection{Theory}\label{sec:theory}
\begin{figure}[h]
	\centering
	\includegraphics[width=.3\textwidth]{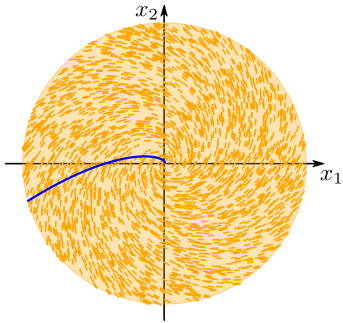}
	\caption{Integral lines in the $(x_1,x_2)$ plane of the director field in \eqref{eq:m_field}. They are logarithmic spirals, which reduce to radii and circles for $\alpha=0$ and $\alpha=\frac{\pi}{2}$, respectively. The blue spiral is of a different nature: it is the reference proto-ridge,  corresponding to the ridge that coincides with the rim of the lamina along the positive $x_1$ axis. The values of the parameters chosen for this illustrative picture are $\alpha=\frac{\pi}{4}$ and $\lambda_1=\sqrt{2/3}$, which according to \eqref{eq:Delta_phi_mu} correspond to $\mu=1.5$.}
	\label{fig:spirals}
\end{figure}

The field $\m$ that we shall consider is described in polar coordinates $(r,\vt)$ by
\begin{equation}
	\label{eq:m_field}
	\m=\cos\alpha\e_r+\sin\alpha\e_\vt,
\end{equation} 
where $\alpha$ is a parameter ranging in $[0,\frac\pi2]$. For $\alpha=0$, $\m$ is the radial field $\e_r$, whereas for $\alpha=\frac{\pi}{2}$ it is the circular field $\e_\vt$ around the center $O$ of the disk. As shown in Fig.~\ref{fig:spirals}, the integral lines of $\m$  are logarithmic spirals winding around $O$, where the director field has a point defect.
To prevent all these lines from winding indefinitely around $O$, we shall remove an arbitrarily tiny disk of radius $r_0\ll R$ centered at $O$, so that our domain becomes a punctured disk free from defects.

Letting $\lambda_1<1$ in \eqref{eq:C_stationary}, the isometric immersion described by $\C_0$ would induce material shrinking along $\m$ and material dilation along $\mper$, whereas the converse would take place for $\lambda_1>1$. Since varying $\alpha\in[0,\frac{\pi}{2}]$  brings $\m$ from the radial to the circumferential orientation, it suffices to choose either $\lambda_1<1$ or $\lambda_1>1$ and change $\alpha$ to go from shrinking radially (and dilating circumferentially) to shrinking circumferentially (and dilating radially). Thus, there is no loss of generality in taking $\lambda_1<1$, as done henceforth for definiteness.\footnote{This heuristic reasoning will be made formal shortly below.} 

Standard computations show that 
\begin{equation}
	\label{eq:m_per_field}
	\mper=-\sin\alpha\e_r+\cos\alpha\e_\vt,
\end{equation}
so that 
\begin{equation}
	\label{eq:various_consequences}
	\cv=\frac1r\e_\vt,\quad c_1=\frac1r\sin\alpha,\quad c_2=\frac1r\cos\alpha,\quad\text{and}\quad\nabla\cv=-\frac{1}{r^2}(\e_r\otimes\e_\vt+\e_\vt\otimes\e_r).
\end{equation}
It readily follows from \eqref{eq:various_consequences}, \eqref{eq:theorema_egregium}, and \eqref{eq:connector_components} that $K\equiv0$, so that the first requirement for the applicability of our method is fulfilled (see also equation (3.6) of \cite{mostajeran:encoding}).

As for the second, we need to find a family of curves that solve \eqref{eq:protoridge_equation} and can be used to decompose $S$. Letting $\lambda_2=1/\lambda_1$ and using \eqref{eq:e_dot_c} in \eqref{eq:protoridge_equation}, we easily see that $\vp\equiv\vp_0$ (so that $\kappa\equiv\e\cdot\cv$) solves \eqref{eq:protoridge_equation}, provided that $t_0:=\tan\vp_0$ is a root of the polynomial equation
\begin{equation}
	\label{eq:polynomial}
	\frac{1}{\lambda_1^4}(\cos\alpha) t^3+(\sin\alpha) t^2+(\cos\alpha) t+\lambda_1^4\sin\alpha=0.
\end{equation}
The only real root of \eqref{eq:polynomial} is
\begin{equation}
	\label{eq:root}
	t_0=-\lambda_1^4\tan\alpha.
\end{equation}

For $\lambda_1<1$, the (constant) angle $\vp_0+\alpha$ that the unit tangent $\e$ to the generic proto-ridge makes with $\e_r$ is positive, but less than $\alpha$,
\begin{equation}
	\label{eq:e_new_representation}
	\e=\cos(\vp_0+\alpha)\e_r+\sin(\vp_0+\alpha)\e_\vt,\quad\vp_0=-\arctan(\lambda_1^4\tan\alpha).
\end{equation}
Thus, proto-ridges too are logarithmic spirals winding in the same sense as the integral lines of $\m$, while lagging behind them (see Fig.~\ref{fig:spirals}). For $\alpha=0$, proto-ridges are radii of the disk, whereas for $\alpha=\frac{\pi}{2}$ they are its inner circles.\footnote{It might be noted that, since $c_1=0$ for $\alpha=0$, the former case corresponds to a rare occasion when \eqref{eq:phi_0} represents indeed a straight segment.}

One proto-ridge will play a special role here. We imagine that the disk sits in the $(x_1,x_2)$ plane, which represents a table supporting a rigid lamina situated in the $(x_1,x_3)$ plane. Upon activation, the disk can freely climb out of the $(x_1,x_2)$ plane, but only in the $x_3\geq0$ half-space; the rim of the lamina sitting along the $x_1$ axis constrains two ridges of $\surface$.\footnote{A tiny gap must be present between the rigid lamina and the elastic disk to allow the latter to glide freely in the $(x_1,x_2)$ plane.} The pre-image of the ridge along $\e_1$ (for $x_1\geq0$) is the \emph{reference} proto-ridge (see again Fig.~\ref{fig:spirals}). All other proto-ridges can be obtained by rotating the reference proto-ridge around the center $O$ of the disk.

We now compute both the length $L$ of each proto-ridge and the length $\ell$ of all ridges. To this end, we first represent the reference proto-ridge through a position vector $\p$ issued from $O$,
\begin{equation}
	\label{eq:reference_proto-ridge}
	\p(\vt)=\varrho(\vt)\e_r.
\end{equation}
To determine the function $\varrho$, we remark that 
\begin{equation}
	\label{eq:remark}
	\e=\frac{\dot{\p}}{|\dot{\p}|},
\end{equation}
where a superimposed dot denotes differentiation with respect to $\vt$. By use of \eqref{eq:e_new_representation},  \eqref{eq:remark} becomes a differential equation for $\varrho$, whose solution is 
\begin{equation}
	\label{eq:rho_solution}
	\varrho(\vt)=r_0\exp\Big({\cot(\vp_0+\alpha)\vt}\Big)\quad\text{for}\quad0\leq\vt\leq\vt_0,
\end{equation}
where
\begin{equation}
	\label{eq:theta_0}
	\vt_0:=\tan(\vp_0+\alpha)\ln\left(\frac{R}{r_0}\right).
\end{equation}

Thus, we easily obtain that 
\begin{equation}
	\label{eq:L}
	L=\int_0^{\vt_0}\sqrt{\varrho^2+\dot{\varrho}^2}\dd\vt=\sqrt{1+\tan^2(\vp_0+\alpha)}(R-r_0).
\end{equation}
On the other hand, since $\vp_0$ is constant, it follows from \eqref{eq:e_representation} that 
\begin{equation}\label{eq:ell}
\ell=(\e\cdot\C_0\e)^{1/2}=L\left(\lambda_1^2\cos^2\vp_0+\frac{1}{\lambda_1^2}\sin^2\vp_0\right)^{1/2},
\end{equation}
which, by \eqref{eq:root} and \eqref{eq:L}, becomes
\begin{equation}
	\label{eq:ell_explicit}
\ell=\frac{\lambda_1(R-r_0)}{\sqrt{\cos^2\alpha+\lambda_1^4\sin^2\alpha}}.
\end{equation}
We see from \eqref{eq:ell_explicit} that  the surface $\surface$ composed of all lifted planar sectors $\surface_j$ is a conical surface with all radii of equal length $\ell$, ranging from $(R-r_0)\lambda_1$ to $(R-r_0)/\lambda_1$ as $\alpha$ varies in $[0,\frac{\pi}{2}]$.

\begin{figure}[h]
	\centering
	\includegraphics[width=.5\textwidth]{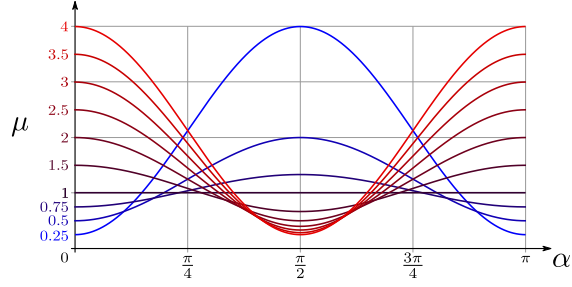}
	\caption{The graphs of $\mu$ against $\alpha$ according to \eqref{eq:Delta_phi_mu} for $\frac12\leq\lambda_1\leq2$. The values of $\mu$ for $\alpha=0$ shown in the picture correspond to $1/\lambda^2_1$.}
	\label{fig:mu}
\end{figure}

To compute the total angle $\Delta$ spanned by the sectors $\surface_j$, we need only recall that the rim of $S$ is mapped into the rim of $\surface$,\footnote{By symmetry, indeed all inner circles of $S$ are mapped into deflated images of the rim of $\surface$.} and so 
\begin{equation}
	\label{eq:metric_identity}
	(\e_\vt\cdot\C_0\e_\vt)^{1/2}=\frac{\ell\Delta}{2\pi(R-r_0)},
\end{equation}
whence it easily follows that 
\begin{equation}
	\label{eq:Delta_phi_mu}
	\Delta=2\pi\mu\quad\text{with}\quad\mu:=\lambda_1^2\sin^2\alpha+\frac{1}{\lambda_1^2}\cos^2\alpha.
\end{equation}
As apparent from \eqref{eq:Delta_phi_mu}, $\mu$ is the effective activation parameter in this illustration of our theory. It depends on $\alpha$ and ranges from $1/\lambda_1^2$ to $\lambda_1^2$ as $\alpha$ varies in $[0,\frac{\pi}{2}]$. Moreover, it is invariant under the transformation that produces the simultaneous changes $\lambda_1\mapsto1/\lambda_1$ and $\alpha\mapsto\frac{\pi}{2}-\alpha$, This shows that taking $\lambda_1<1$ and letting $\alpha$ cover the whole interval $[0,\frac{\pi}{2}]$ (as done here)  effectively subsumes the case $\lambda_1>1$.\footnote{This formally recaps the heuristic argument given above.} In the whole excursion of $\alpha$, $\mu$ ranges from above to below unity. Figure~\ref{fig:mu} depicts the graph of $\mu$ against $\alpha$ drawn for several values of $\lambda_1$.

For $\mu<1$, the total angle deficiency (to $2\pi$) would make $\surface$ shrink into a circular cone, if it were not impeded by the lamina to so. If we wish to wrap the disk around the lamina to grab it, remaining in the class of deformations described in Sec.~\ref{sec:strategy}, we must take $\mu>1$, as in this case the total angle excess induces  folds in $\surface$.

The optimal shape of $\surface$ subject to the physical constraints imposed by both table and lamina will be determined in Sec.~\ref{sec:numerics} by minimizing $\free_r$ in \eqref{eq:total_ridge_energy}, which here reduces to
\begin{equation}
	\label{eq:total_ridge_energy_reduced}
	\free_r=\ell\sum_{j=1}^Nf_r=\frac{R-r_0}{\sqrt{\mu}}\sum_{j=1}^Nf_r,
\end{equation}
where $N$ is the (chosen) total number of ridges and $f_r$ is as in \eqref{eq:ridge_energy_density}. To fully express this latter in the present setting, we still need to compute $(\n\cdot\tangent)$ and $(\nper\cdot\tangent)$. We get the former from
\begin{subequations}\label{eq:n_dot_x}
\begin{equation}
	\label{eq:n_dot_t}
	\n\cdot\tangent=\frac{\m\cdot\C_0\e}{(\m\cdot\C_0\m)^{1/2}(\e\cdot\C_0\e)^{1/2}}=\frac{\lambda_1^2\cos\vp_0}{\sqrt{\sin^2\vp_0+\lambda_1^4\cos^2\vp_0}}=\frac{\cos\alpha}{\lambda_1\sqrt{\mu}},
\end{equation}
which follows from \eqref{eq:e_representation} and \eqref{eq:root}. Then from \eqref{eq:n_dot_t} we get
\begin{equation}
	\label{eq:n_dot_t_perp}
	\n\cdot\tangent_\perp=\frac{\lambda_1\sin\alpha}{\sqrt{\mu}}.
\end{equation}
\end{subequations}
Equations \eqref{eq:n_dot_x} show that the integral lines of $\n$ are conical spiral making a constant angle with the ridges of $\surface$.

Finally, \eqref{eq:total_ridge_energy_reduced} expands into
\begin{equation}
	\label{eq:total_ridge_energy_expanded}
	\free_r=\frac83h^2\frac{R-r_0}{\sqrt{\mu}}\left(1-\frac{s}{s+1}\frac{1}{\lambda_1^2\mu^2}\cos^2\alpha\sin^2\alpha\right)\sum_{j=1}^N\arccos^2(\normal_j\cdot\normal_{j+1}),
\end{equation}
where $\normal_j$ is the unit normal to $\surface_j$, with the identification $\normal_{N+1}=\normal_1$. The complete formula for $\free_r$ in \eqref{eq:total_ridge_energy_expanded} is needed only to compare minimal energies corresponding to different values of $\alpha$; for a given $\alpha$, the optimal shape of $\surface$ can be determined by omitting the inessential prefactor of the sum, but $\mu$ (and so $\alpha$) will still come into play through the requirement on the total angle $\Delta$ in \eqref{eq:Delta_phi_mu}, which will act as a further  constraint for $\free_r$, of a mathematical origin.

\subsection{Illustration}\label{sec:numerics}
Hereafter we shall scale lengths to $\ell$ and reduce $\free_r$ in \eqref{eq:total_ridge_energy_expanded} to its dimensionless counterpart,
\begin{equation}\label{eq:total_ridge_energy_reduced_and_scaled}
	F_r:=\sum_{j=1}^N\arccos^2(\normal_j\cdot\normal_{j+1}),
\end{equation}
subject to \eqref{eq:Delta_phi_mu} and to the conditions that require all the ridges to live in the half-space $x_3\geq0$ and prohibit both self-intersection and crossing of the lamina in the $(x_1,x_3)$ plane.

The minimization of $F_r$ under these constraints was performed numerically with the algorithm outlined in Appendix~\ref{sec:numerical}. The physical constraints offered by lamina and table were enforced by penalizing both the half-space $x_3<0$ and the plane $x_2=0$ with localized highly repulsive potentials. By symmetry, at least for moderate values of $\mu$, self-intersection could only take place on the  plane  $x_1=0$. There, to prevent self-intersection, we placed a virtual lamina, which we treated precisely as the real one. More algorithmic details are given in Appendix~\ref{sec:numerical}; here, we report and comment upon the outcomes of our computations.

\begin{figure}
	\centering
	\begin{subfigure}[t]{.48\textwidth}
		\centering
		\includegraphics[width=1\textwidth]{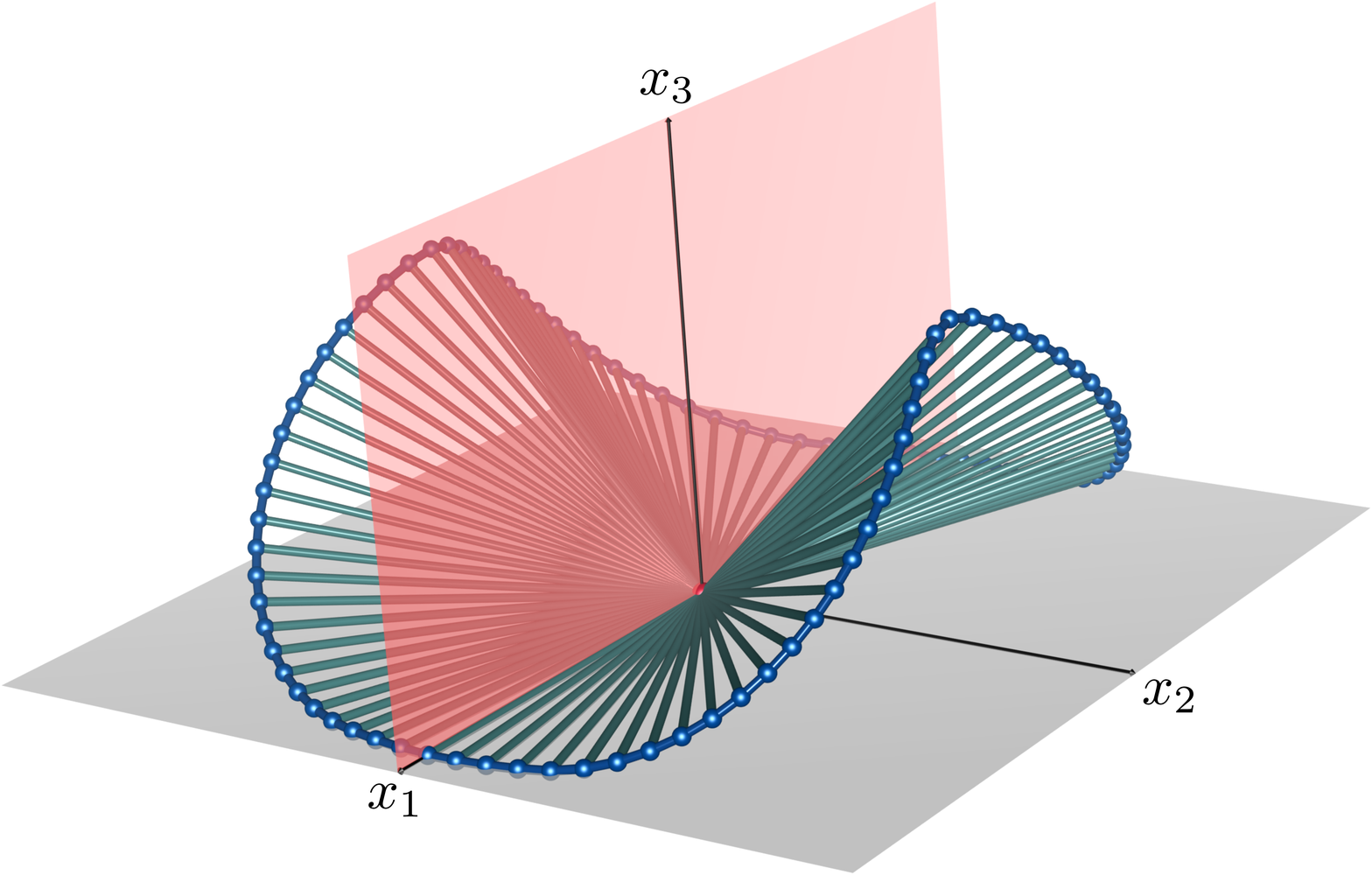}
		\caption{$\mu=1.10$}
	\end{subfigure}
	\begin{subfigure}[t]{.48\textwidth}
		\centering
		\includegraphics[width=1\textwidth]{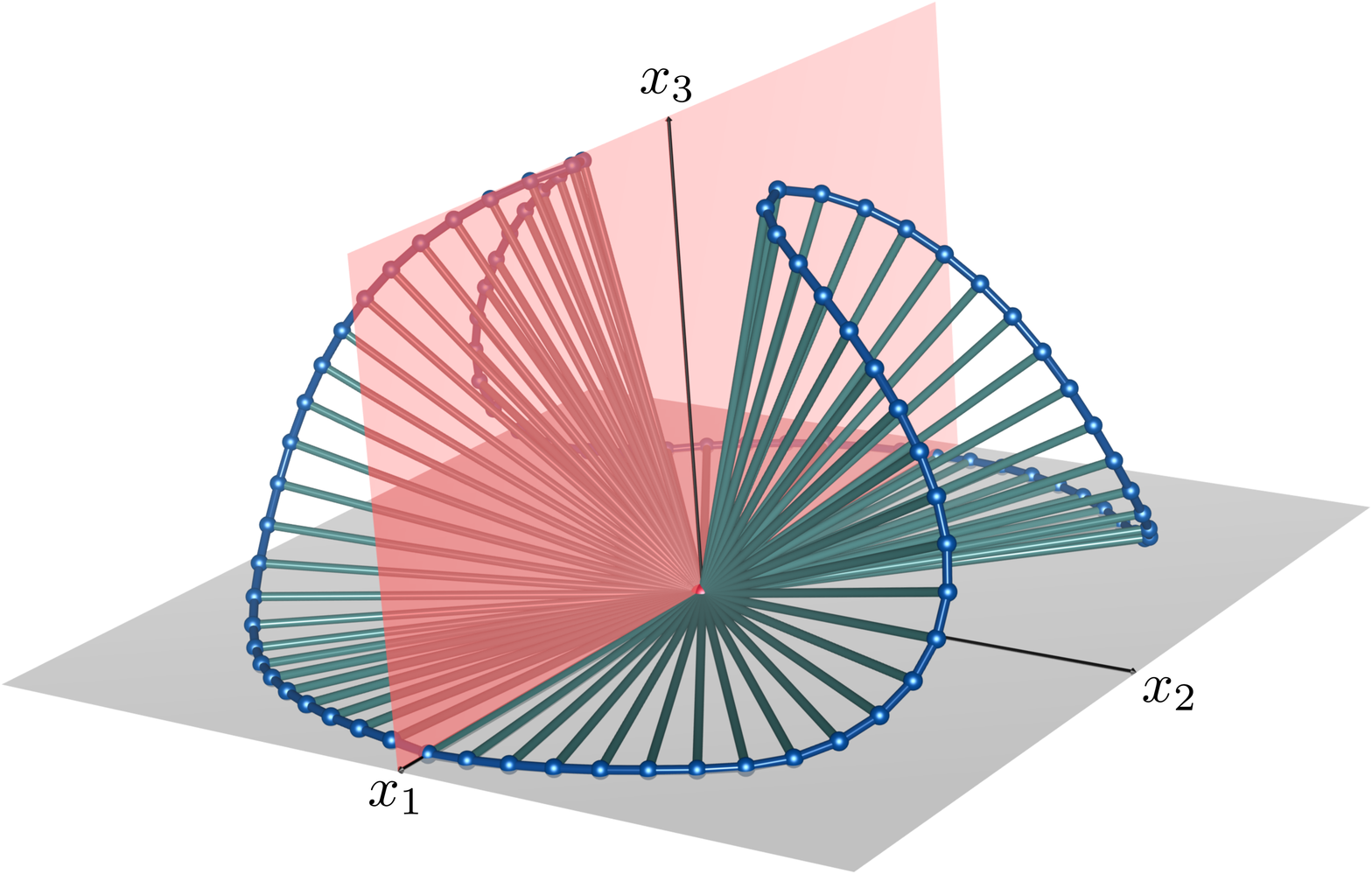}
		\caption{$\mu=1.50$}
	\end{subfigure}\quad
	\begin{subfigure}[t]{.48\textwidth}
		\centering
		\includegraphics[width=1\textwidth]{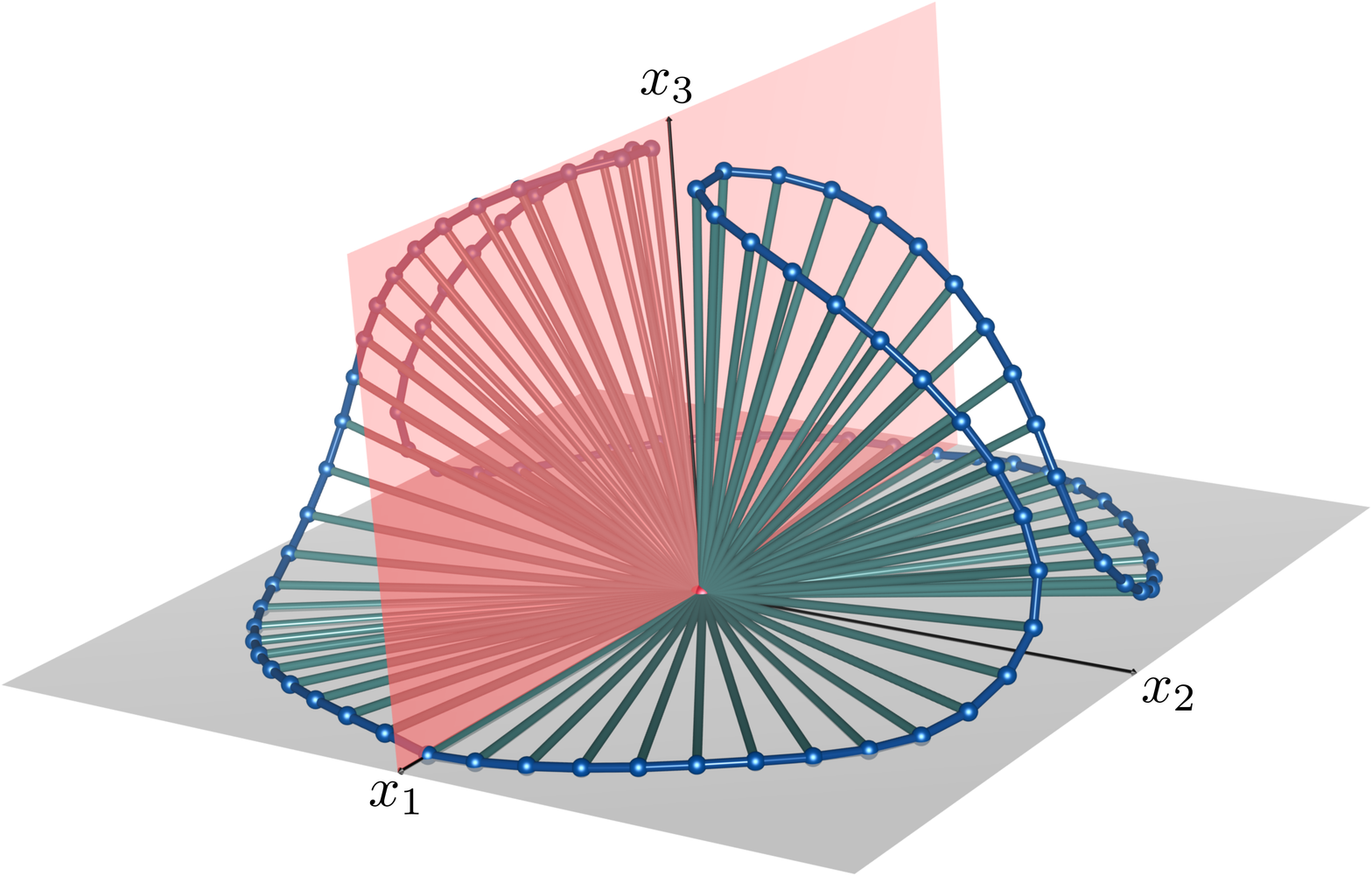}
		\caption{$\mu=1.80$}
	\end{subfigure}\quad
	\begin{subfigure}[t]{.48\textwidth}
		\centering
		\includegraphics[width=1\textwidth]{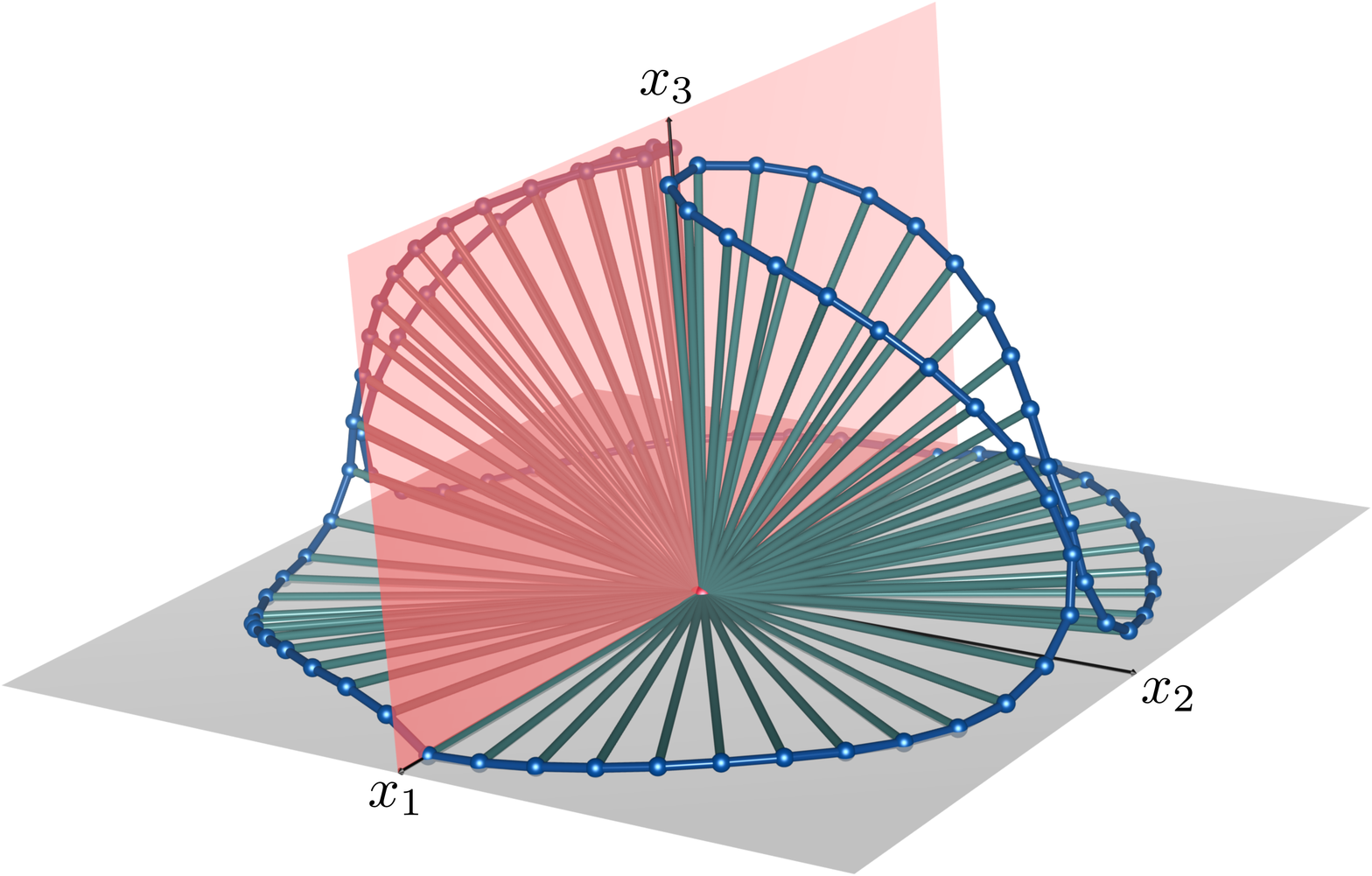}
		\caption{$\mu=1.95$}
	\end{subfigure}\quad
	\begin{subfigure}[t]{.48\textwidth}
		\centering
		\includegraphics[width=1\textwidth]{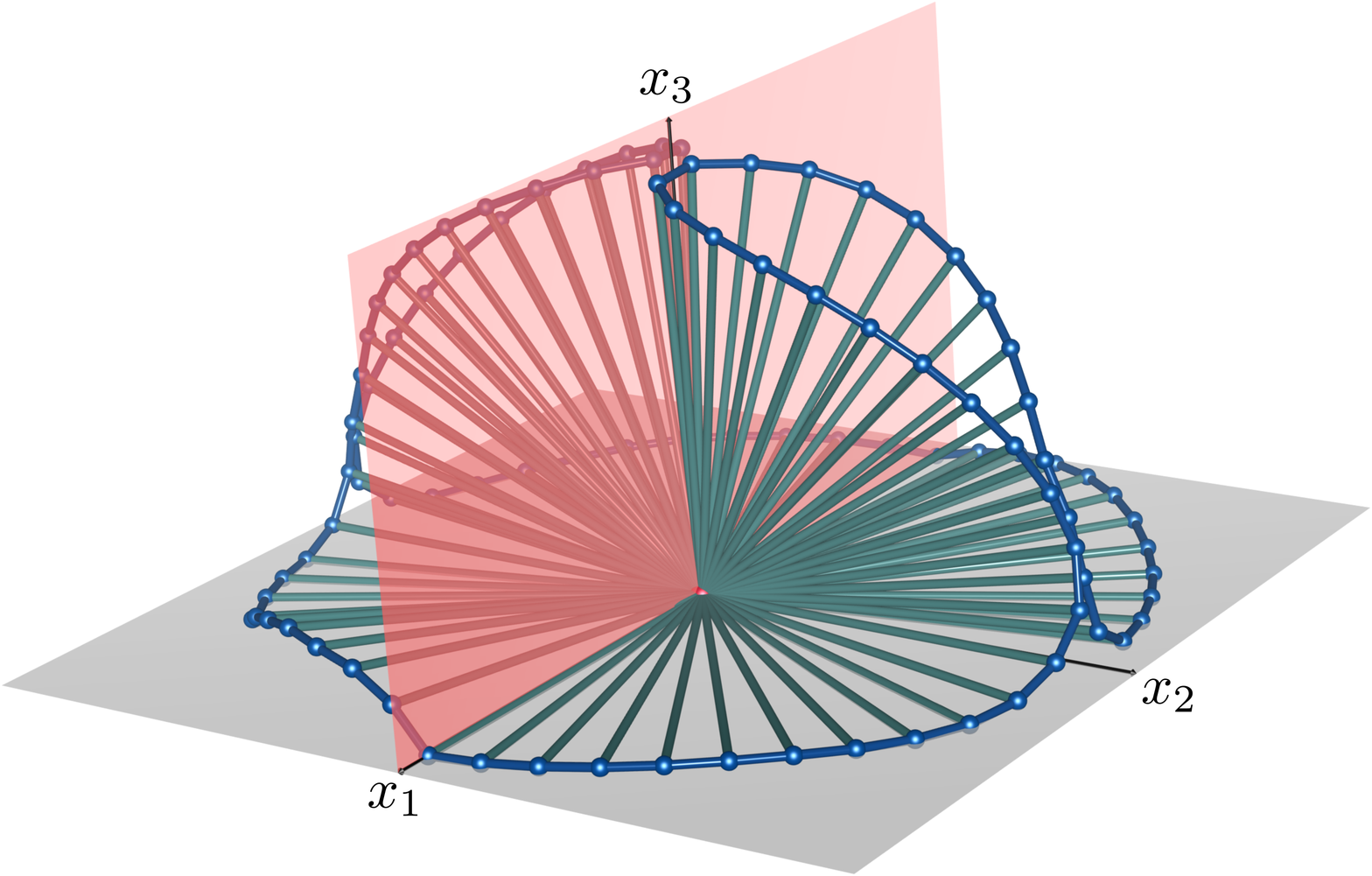}
		\caption{$\mu=2.00$}
	\end{subfigure}\quad
	\begin{subfigure}[t]{.48\textwidth}
		\centering
		\includegraphics[width=1\textwidth]{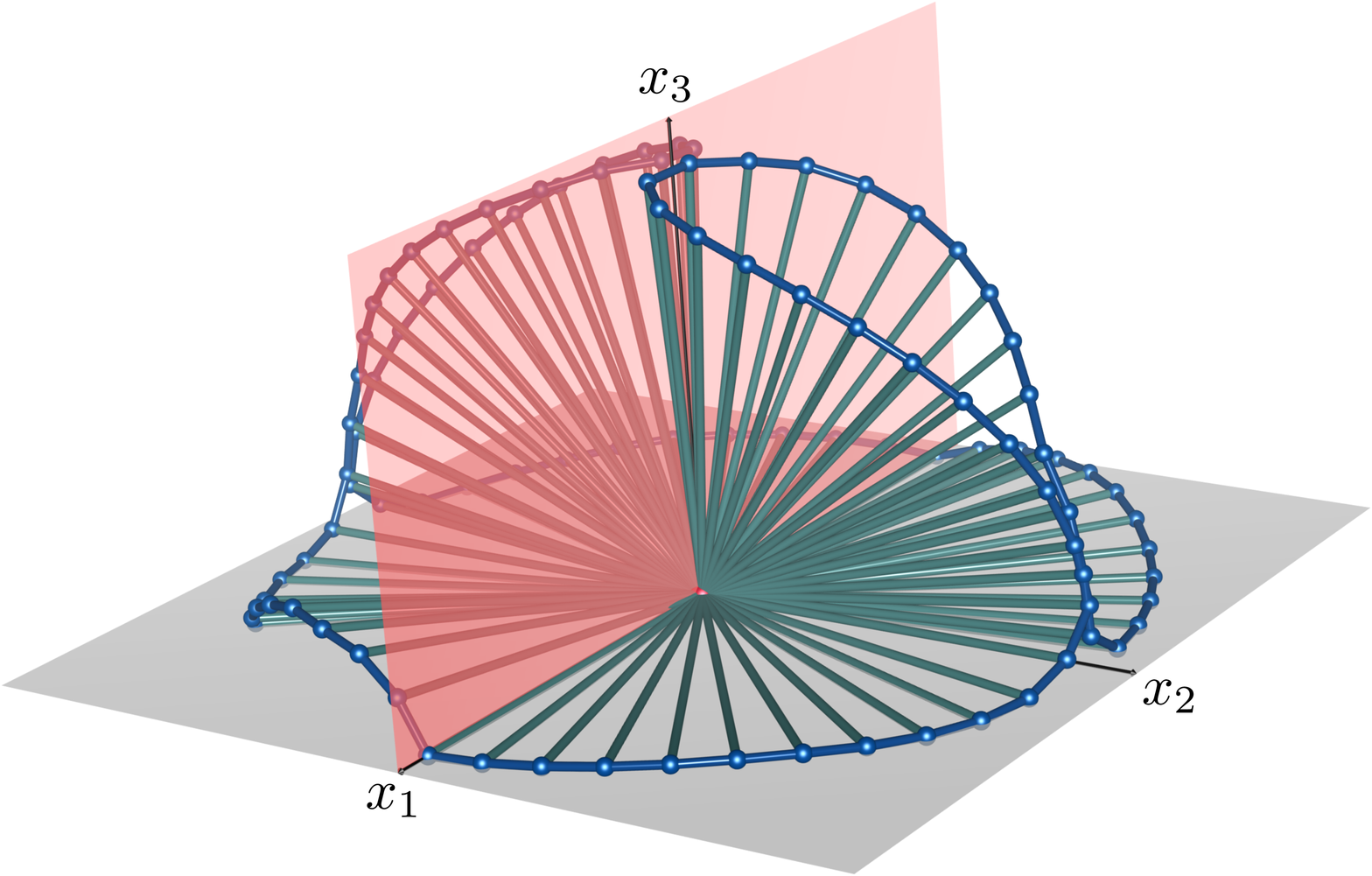}
		\caption{$\mu=2.05$}
	\end{subfigure}
	\caption{A gallery of equilibrium shapes of the activated disk acquired for increasing values of the effective activation parameter $\mu$ and total number of ridges $N=96$. The lamina is the reddish $(x_1,x_3)$ plane, while the table is the grayish $(x_1,x_2)$ plane. The red dot is the center $O$ of the undeformed disk (and the apex of the conical activated shapes). Ridges have the same length in all panels, as we have scaled lengths to $\ell=(R-r_0)/\sqrt{\mu}$. In reality, for increasing $\mu$, they shrink accordingly. All pictures were drawn using  VESTA software~\cite{momma:vesta}.}
	\label{fig:finger_gallery}
\end{figure}

The galley of images in Fig.~\ref{fig:finger_gallery} reproduce the energy minimizing shapes of the activated disk for $1.10\leq\mu\leq2.05$. Upon increasing $\mu$, the deformed shape soon develops a pair of \emph{fingers} spurring out of a \emph{palm} flattened by the rim's lamina. At $\mu\approx1.80$, while the fingers raise in a grabbing posture, tiny waves develop on the palm to release elastic energy, in an attempt to develop further fingers to secure, as it were, a firmer grip. At $\mu\approx1.95$, self-contact takes place in the two prominent fingers and the equilibrium shape flattens on the virtual lamina. Symmetry was only invoked to locate the latter, not enforced on the class of admissible shapes. The symmetry exhibits in Fig.~\ref{fig:finger_gallery} is a genuine outcome of the minimization process. 

In our model, the lamina is just a plane, with zero thickness. We stopped our search at $\mu=2.05$, when we reckoned that the tips of the grabbing fingers were so close that they should have already come in contact with the real body. 

\section{Conclusions}\label{sec:conclusions}
Actuating a thin NPN sheet  entails inducing a given metric tensor on an originally flat surface. In general, the resulting deformation in three dimensions carries a predominant stretching energy, accompanied by a smaller bending component. The search for isometric immersions has so far been the most common tool to explore the variety of activable shapes. No role is played by the bending energy in this approach, despite the fact that when isometric immersions superabound the bending energy appears to be the appropriate selection criterion.

This paper took the avenue of computing the minimum of the bending energy over a class of piece-wise smooth isometries, the ridged isometries, for which the bending energy is concentrated on straight ridges in the actuated surface and scales quadratically in the sheet's thickness.

Our minimization strategy relied on two requirements: 
\begin{inparaenum}[(1)]
\item the ability to impose the desired metric on a planar deformations and
\item the possibility of decomposing the original flat surface in sectors delimited by proto-ridges, pre-images of straight ridges.
\end{inparaenum} 
Satisfying both requirements ensures that the deformed planar sectors, although not all fitting anymore, can be lifted in space so as to minimize the total ridge energy. 

We gave necessary and sufficient conditions for meeting both requirements and making our strategy viable in general. The first condition demands that the Gaussian curvature associated with the prescribed metric vanishes, which makes the activable surface necessary developable. The second condition is a differential equation whose solutions describe the proto-ridges emanating from a given hub; the existence of proto-ridges was thus established in general, but it remains to be seen whether they can reach the boundary of a particular domain. 

Armed with these (more or less) explicit conditions for the applicability of our approximation scheme, we illustrated our theory by studying a problem with potential practical application. We asked whether a disk with an imprinted (spiraling) hedgehog could be used to grab a rigid lamina surmounting it. We found numerically the optimal shape of the deformed disk and determined  values of the effective activation parameter $\mu$ that would bring the folds of the deformed disk close enough to the sides of  the lamina  to grab it.

We see (at least) two possible extensions of our work. For the specific grabbing problem, it would perhaps be interesting to find the values of $\mu$ that achieve the desired goal for other, possibly less symmetric shapes of the object to grab.

More generally, it would be desirable to apply the ridge approximation proposed here to a number of other imprinted director fields $\m$, for which the Gaussian curvature $K$ vanishes. The PDE that identifies $\m$  for prescribed $K$ has already been introduced (see, e.g., (2.3) of \cite{mostajeran:encoding}) and studied in special classes \cite{mostajeran:encoding,aharoni:geometry}. To our knowledge, the most general family of solutions for $K\equiv0$ has not yet been characterized. Applying our method to that class might enlarge significantly the family of developable surfaces that can be activated. 

\section*{Data Availability Statement}
The numerical data that support the findings in Sec.~\ref{sec:numerics} and Appendix~\ref{sec:numerical} are available from the corresponding author upon reasonable request.

\begin{acknowledgments}
	The work of A.P. was supported financially by the Department of Mathematics of the University of Pavia as part of the activities funded by the Italian MIUR under the nationwide Program ``Dipartimenti di Eccellenza (2018-2022).''
\end{acknowledgments}

\appendix
\section{Two Mathematical Tasks}\label{sec:tasks}
Here we collect the mathematical details needed to accomplish the tasks we assigned to ourselves in Sec.~\ref{sec:strategy} to check the viability of our strategy.

\subsection{First Task}\label{sec:first_task}
For a generic $S_j$, let $\F$ be the gradient $\nabla\y_j$ of $\y_j$. By the polar decomposition theorem (see, for example \cite[p.\,31]{gurtin:mechanics}),
\begin{equation}
	\label{eq:polar_decomposition}
	\F=\R\U,
\end{equation}
where $\R$ is a rotation by angle $\chi$ around $\e_3$ and $\U$ is required to be $\sqrt{\C_0}$.\footnote{Here we take both $\lambda_1$ and $\lambda_2$ in \eqref{eq:C_stationary} as constants, but not necessarily such that $\lambda_1\lambda_2=1$. The latter constraint can easily be enforced whenever needed.} Since 
\begin{equation}
	\label{eq:rotation_representation}
	\R=\cos\chi\I+\sin\chi\W_3,
\end{equation}
where $\I$ is the (two-dimensional) identity and $\W_3$ is the skew-symmetric tensor associated with $\e_3$, it follows from \eqref{eq:C_stationary} that 
\begin{equation}
	\label{eq:F_representation}
	\F=\cos\chi(\lambda_1\m\otimes\m+\lambda_2\mper\otimes\mper)+\sin\chi(\lambda_1\mper\otimes\m-\lambda_2\m\otimes\mper).
\end{equation}

With a prescribed stretching tensor $\U$, there is no guarantee that $\F$ in \eqref{eq:F_representation} be integrable. The integrability condition is the symmetry (in the last two legs) of the third-order tensor $\nabla\F$, which amounts to the (ordinary) symmetry of the two second-order tensors obtained by contraction of the first leg of $\nabla\F$ with the vectors of the basis $(\m,\mper)$.

By use of the following equations, which characterize the connector $\cv$ (see also \cite{ozenda:blend}),
\begin{equation}
	\label{eq:connector_equations}
	\nabla\m=\mper\otimes\cv,\quad\nabla\mper=-\m\otimes\cv,\quad\text{and}\quad\nabla\cv-(\nabla\cv)\trans=\zero,
\end{equation}
we arrive at
\begin{subequations}
	\label{eq:nabla_F_components}
	\begin{gather}
		\m\cdot\nabla\F=-\lambda_1\sin\chi\m\otimes\nabla\chi-\lambda_2\cos\chi\mper\otimes\nabla\chi+(\lambda_1-\lambda_2)(\cos\chi\mper\otimes\cv-\sin\chi\m\otimes\cv),\\
		\mper\cdot\nabla\F=-\lambda_2\sin\chi\mper\otimes\nabla\chi+\lambda_1\cos\chi\otimes\nabla\chi+(\lambda_1-\lambda_2)(\cos\chi\m\otimes\cv+\sin\chi\mper\otimes\cv).
	\end{gather}
\end{subequations}
It is now a simple matter to show that the requirement of symmetry for both these tensors reduces to a linear system for the components of $\nabla\chi$ in the local basis $(\m,\mper)$, whose unique solution delivers
\begin{equation}
	\label{eq:nabla_theta}
	\nabla\chi=\frac{\lambda_1-\lambda_2}{\lambda_1\lambda_2}(\lambda_1c_1\m-\lambda_2c_2\mper),
\end{equation}
where use has been made of \eqref{eq:connector_components} for the definition of the components of $\cv$.

Thus, whenever \eqref{eq:nabla_theta} is valid, $\F$ is integrable, so that the integrability of $\F$ follows from the integrability of $\nabla\chi$. The latter condition is equivalent to the symmetry of $\nabla^2\chi$. By \eqref{eq:connector_equations} and \eqref{eq:nabla_theta}, the latter tensor can be written as
\begin{equation}
	\label{eq:hessian_theta}
	\nabla^2\chi=\frac{\lambda_1-\lambda_2}{\lambda_1\lambda_2}(\lambda_1c_1\mper\otimes\cv+\lambda_1\m\otimes\nabla c_1+\lambda_2c_2\m\otimes\cv-\lambda_2\mper\otimes\nabla c_2),
\end{equation}
which is symmetric if, and only if,
\begin{equation}
	\label{eq:hessian_theta_symmetry_condition}
	\lambda_1\nabla c_1\m+\lambda_2\nabla c_2\otimes\m=\lambda_1c_1^2-\lambda_2c_2^2.
\end{equation}
Since, again by \eqref{eq:connector_equations}, 
\begin{equation}
	\label{eq:nabla_c_1_2}
	\nabla c_1=(\nabla\cv)\m+c_2\cv\quad\text{and}\quad\nabla c_2=(\nabla\cv)\mper-c_1\cv,
\end{equation}
equation \eqref{eq:hessian_theta_symmetry_condition} is equivalent to 
\begin{equation}
	\label{eq:hessian_symmetry_equivalent}
	\frac{\lambda_1^2-\lambda_2^2}{\lambda_1\lambda_2}(c_2^2-c_1^2+c_{12})=0,
\end{equation}
which by \eqref{eq:theorema_egregium} reduces to $K=0$ when also \eqref{eq:det_C_constrained} is enforced.\footnote{It is perhaps worth noting in passing that when $\lambda_1=\lambda_2$ equation \eqref{eq:hessian_symmetry_equivalent} is identically satisfied for all $\m$'s and, correspondingly, by \eqref{eq:nabla_theta} $\chi$ is any constant.}

This is our first desired conclusion: whenever $\m$ is such that \eqref{eq:hessian_symmetry_equivalent} is satisfied, each $S_j$ can be immersed isometrically in the plane with $\C=\C_0$ and a rotation field $\chi$ obtained by integrating \eqref{eq:nabla_theta}. Clearly, $\chi$ may suffer (constant) jumps along the arcs separating adjacent $S_j$'s, so as to scatter all $\surface_j$'s on the plane. But recombining them in a single planar $\surface$ would be impossible (uless $\lambda_1=\lambda_2$), as shown in Fig.~\ref{fig:decomposition} of the main text. 

\subsection{Second Task}\label{sec:second_task}
Let $\e$ be the unit tangent vector to a proto-ridge $C$. The unit tangent $\tangent$ to the ridge $\curve$ is given by
\begin{equation}
	\label{eq:tangent_equation}
	\tangent=\frac{\F\e}{|\F\e|}.
\end{equation}
Denoting by a prime $'$ the differentiation with respect to the arc length on $C$, we require that $\tangent'=\zero$. It follows from \eqref{eq:tangent_equation} that 
\begin{equation}
	\label{eq:tangent_drivative}
	\tangent'=\frac{1}{|\F\e|}\bigg((\F\e)'-\Big((\F\e)'\cdot\tangent\Big)\tangent\bigg),
\end{equation}
so that our desired requirement reduces to $(\F\e)'\parallel\tangent$, that is, to
\begin{equation}
	\label{eq:parallelism_condition}
	(\F\e)'\times\F\e=\zero.
\end{equation}

Now,
\begin{equation}
	\label{eq:preliminary_equation}
	(\F\e)'=\nabla\F\cdot\e\otimes\e+\F\e',
\end{equation}
where the two last legs of $\nabla\F$ are saturated with the components of $\e$. To expand equation \eqref{eq:preliminary_equation}, it is convenient to represent $\e$ as in \eqref{eq:e_representation}, so that, in particular,
\begin{equation}
	\label{eq:e_dot_c}
	\e\cdot\cv=c_1\cos\vp+c_2\sin\vp.
\end{equation} 
It readily follows from \eqref{eq:e_representation} that 
\begin{equation}
	\label{eq:e_prime}
	\e'=(\e\cdot\cv+\vp')\e_\perp,
\end{equation}
where $\e_\perp:=\e_3\times\e$, so that $\kappa:=\e\cdot\cv+\vp'$ is the (signed) curvature of $C$.

With little more labor, it follows from \eqref{eq:e_representation} and \eqref{eq:nabla_F_components} that 
\begin{subequations}
\label{eq:preliminary_equation_1}
\begin{equation}
	\begin{split}
	\nabla\F\cdot\e\otimes\e&=(\nabla\chi\cdot\e)[(\lambda_1\sin\chi\cos\vp+\lambda_2\cos\chi\sin\vp)\m+(\lambda_1\cos\chi\cos\vp-\lambda_2\sin\chi\sin\vp)\mper]\\
	&+(\lambda_1-\lambda_2)(\e\cdot\cv)[\sin(\vp-\chi)\m+\cos(\vp-\chi)\mper].	
	\end{split}	
\end{equation}
We similarly obtain that
\begin{gather}
	\F\e'=\kappa[(-\lambda_1\cos\chi\sin\vp+\lambda_2\sin\chi\cos\vp)\m+(\lambda_2\cos\chi\cos\vp-\lambda_1\sin\chi\sin\vp)\mper],\\
	\F\e=(\lambda_1\cos\chi\cos\vp-\lambda_2\sin\chi\sin\vp)\m+(\lambda_2\cos\chi\sin\vp+\lambda_1\sin\chi\cos\vp)\mper.
\end{gather}
\end{subequations}
It is now a tedious, but simple task to show that \eqref{eq:parallelism_condition} reduces to \eqref{eq:protoridge_equation} in the main text.  

\section{Numerical Algorithm}\label{sec:numerical}
Since the effect of both $\alpha$ and $\lambda_1$ on $F_r$ is mediated by $\mu$, different choices of $\alpha$ and $\lambda_1$ that produce the same value of $\mu$ will lead to the same optimal shape of $\surface$; for this reason, the numerical optimization we are discussing here will depend only on the parameter $\mu$ and the chosen number of ridges $N$. For such fixed parameters, all the $\surface_j$'s are supposed to be identical planar circular sectors  with amplitude (inter-ridge angle) $\frac{2\pi}{N}\mu$.
\begin{figure}[h]
	\centering
	\includegraphics[width=.6\textwidth]{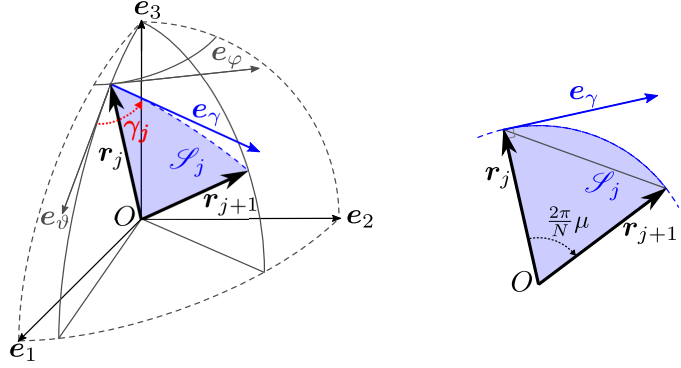}
	\caption{Geometrical meaning of angles $\gamma_j$, used as variables in the minimization process of \eqref{eq:weighted_energy}. Each $\gamma_j$ represents the inclination of the circular sector $\surface_j$ on the  plane through $\e_3$ containing $\ridge_j$. If $\ridge_j$ is known, $\ridge_{j+1}$ is determined by $\gamma_j$ since the amplitude $\frac{2\pi\mu}{N}$ of $\surface_j$ is fixed. Here $(\ridge_j, \e_\vartheta, \e_\varphi)$ is the local basis at the tip of $\ridge_j$, while $\e_\gamma$ is tangent to the arc of $\surface_j$, and $\ridge_{j+1}$ in given by \eqref{eq:next_ridge}.}
	\label{fig:gamma}
\end{figure} 

We represent the $j$-th ridge through the unit vector $\ridge_j$ issued from $O$. As shown in Figure \ref{fig:gamma}, the angle $\gamma_j$ represents the inclination\footnote{Angles $\gamma_j$'s play here a role similar to that played   in \cite{pedrini:ridge} by the single angle $\beta$, which designated the inclination on the $(x_1,x_2)$ plane of all $\surface_j$'s, in the very special case where the number $N$ of ridges is also the number of macroscopic folds of $\surface$.}
of $\surface_j$ on the plane through $\e_3$ containing $\ridge_j$. By the knowledge of $\Delta$ in \eqref{eq:Delta_phi_mu} and $\gamma_j$, the ridge $\ridge_{j+1}$ remains completely determined,
\begin{equation}
	\label{eq:next_ridge}
	\ridge_{j+1} =  \cos\frac{2\pi\mu}{N}\ridge_j + \sin\frac{2\pi\mu}{N}\e_\gamma,
\end{equation}
where
\begin{equation}
	\label{eq:e_phi_theta}
	\e_\gamma:=\cos\gamma_j\e_\vartheta + \sin\gamma_j\e_\varphi,\quad
	\e_\varphi=\frac{\e_3\times\ridge_j}{|\e_3\times\ridge_j|},\quad\text{and}\quad \e_\vartheta=\e_\varphi\times\ridge_j.
\end{equation}
Thus, $F_r$ in \eqref{eq:total_ridge_energy_reduced_and_scaled} becomes
\begin{equation}\label{eq:total_ridge_energy_reduced_normals}
	F_r = \sum_{j=1}^N\arccos^2\left( \frac{\ridge_j\times\ridge_{j+1}}{|\ridge_j\times\ridge_{j+1}|}\cdot\frac{\ridge_{j+1}\times\ridge_{j+2}}{|\ridge_{j+1}\times\ridge_{j+2}|}\right) 
\end{equation}
with $\ridge_{1}=\e_1$ and $\gamma_{N+1}=\gamma_{1}$.
%%%%%%%%%%%%%%%

To force the sequence of ridges to loop back on itself while preventing any self-intersection of the deformed disk and any intersection with the external obstacles (the vertical lamina and the horizontal table), we impose the following further geometrical constraints
\begin{gather}
	\ridge_{N+1} = \ridge_1 = \e_1 \quad\text{and}\quad \ridge_{\frac{N}{2}+1} = -\ridge_1 = -\e_1 \label{eq:constr1},\\
	\ridge_j\cdot\e_1 \left\{
	\begin{split}
		&\geq0 \qquad\text{for }\ 1\leq j\leq\frac{N}{4}+1\ \text{ or }\ \frac{3N}{4}+1\leq j\leq N,\\
		&\leq0 \qquad\text{for }\ \frac{N}{4}+1\leq j\leq\frac{3N}{4}+1,
	\end{split}\right. \label{eq:constr2}\\
	\ridge_j\cdot\e_2 \left\{
	\begin{split}
		&\geq0 \quad\text{for }\ 1\leq j\leq\frac{N}{2}, \\
		&\leq0 \quad\text{for }\ \frac{N}{2}+2\leq j\leq N,
	\end{split}\right. \label{eq:constr3},\\
	\ridge_j\cdot\e_3\geq0 \qquad \text{for }\ j=1,\dots,N. \label{eq:constr4}
\end{gather}
These are enforced as weighted penalties to energy \eqref{eq:total_ridge_energy_reduced_normals}:
\begin{equation}\label{eq:weighted_energy}
	\begin{split}
		\mathcal{F}_r &:= \sum_{j=1}^N\arccos^2\left( \frac{\ridge_j\times\ridge_{j+1}}{|\ridge_j\times\ridge_{j+1}|}\cdot\frac{\ridge_{j+1}\times\ridge_{j+2}}{|\ridge_{j+1}\times\ridge_{j+2}|}\right)
		+ w_1\left(|\ridge_{N+1} - \e_1|^2 + |\ridge_{\frac{N}{2}+1} + \e_1|^2 \right) \\
		& + w_2 \left[ \sum_{j=1}^{\frac{N}{4}+1} \max(0, -\ridge_j\cdot\e_1) +  \sum_{j=\frac{3N}{4}+1}^{N} \max(0, -\ridge_j\cdot\e_1) +  \sum_{\frac{N}{4}+1}^{\frac{3N}{4}+1} \max(0, \ridge_j\cdot\e_1)\right] \\
		& + w_3 \left[ \sum_{j=1}^{\frac{N}{2}} \max(0, -\ridge_j\cdot\e_2) +  \sum_{j=\frac{N}{2}+2}^{N} \max(0, \ridge_j\cdot\e_2)\right] 
		+ w_4 \sum_{j=1}^{N} \max(0, -\ridge_j\cdot\e_3)
	\end{split}
\end{equation}
with $w_1, w_2, w_3, w_4\geq0$, weights to be arranged.

While avoiding a gap in the deformed disk by gluing together the first and the last ridge, condition \eqref{eq:constr1} also freezes both the first ridge $\ridge_1$ and the middle ridge $\ridge_{\frac{N}{2}+1}$ along the $x_1$ axis, where the lamina meets the table. On the other hand, conditions \eqref{eq:constr2}--\eqref{eq:constr4} compel each quarter of the deformed disk to stay in the corresponding octant above the $(x_1,x_2)$ plane. More precisely, \eqref{eq:constr2} and \eqref{eq:constr4}  prevent the interpenetration of the disk with the lamina and the table, respectively, while \eqref{eq:constr3} avoids self-intersection of the disk by making the $(x_2,x_3)$ plane  a virtual lamina, impenetrable to all ridges,   but  $\ridge_{\frac{N}{4}+1}$ and $\ridge_{\frac{3N}{4}+1}$, which must glide on it, by symmetry.

We seek the optimal deformed shape $\surface$ by progressively reducing the value of $\mathcal{F}_r$ in \eqref{eq:weighted_energy} via a stochastic gradient descent method with momentum (Adam) \cite{kingma:adam} over the variables $\gamma_j$'s, and by progressively increasing weights $w_1, w_2, w_3, w_4$.

Concerning the initialization of the $\gamma_j$'s which leads to the results presented in Sec.~\ref{sec:numerics}, a preliminary exploration (with $w_2=w_3=w_4=0$, different values of $\mu$ and a plethora of randomly chosen initial angles $\gamma_j$) revealed a symmetrical behavior of the deformed disk, but also suggested the existence of a variety of stationary deformed configurations. We combined the experience acquired in this exploration with the minimal construction presented in \cite{pedrini:ridge} and started from $\mu=\mu_\mathrm{i}=1.1$ with initial angles
\begin{equation}\label{eq:gamma_init}
	\gamma^{(\mathrm{i})}_j := \left\{
	\begin{aligned}
		&2\beta\sqrt{J_j(1-J_j)} &&\quad\text{for }\ 1\leq j \leq\frac{N}{4} \qquad \text{ with }\ J_j:=\frac{j-1}{\frac{N}{4}}, \\
		&-\gamma^{(\mathrm{i})}_{j-\frac{N}{4}} &&\quad\text{for }\ \frac{N}{4} + 1\leq j \leq\frac{N}{2}, \\
		&\gamma^{(\mathrm{i})}_{j-\frac{N}{2}} &&\quad\text{for }\ \frac{N}{2} + 1\leq j \leq N,
	\end{aligned}\right.
\end{equation}
where
\begin{equation}\label{eq:beta}
	\beta:=\arccos\frac{\tan\frac{\pi}{4}}{\tan\left(\frac{\pi}{4}\mu_\mathrm{i}\right)}.
\end{equation}
This initialization\footnote{In the actual process, the initialization was slightly changed to avoid any initial angle to be zero.} almost satisfies \eqref{eq:constr1}--\eqref{eq:constr4} and produces an initial configuration of the deformed disk with two macroscopic folds, mirror-symmetric to one another  with respect to the real lamina and almost symmetrical with respect to the virtual lamina. Each fold  lifts off the $(x_1,x_2)$ plane, approaching  the lamina, and smoothly reaches its peak almost exactly on the $(x_2,x_3)$ plane. Such symmetry properties are not imposed during the optimization process, but they are spontaneously acquired and preserved up to the final optimal configuration.
After this first optimization, we progressively increased $\mu$, by using the optimal configuration found at each step as initialization in the optimization process for the next value of $\mu$.

The implementation of the described algorithm was done in Python 3, by making use of the automatic differentiation provided by TensorFlow \cite{abadi:tensorflow}, a free and open-source symbolic math library widely used in Machine Learning. To speed up computations, we ran all numerical experiments on Google Colaboratory.

%\nocite{*}
\bibliography{elastomers}

\end{document}